\documentclass[12pt]{article}
\usepackage{epsf}

\begin{document}


\newcommand\balpha{\mbox{\boldmath $\alpha$}}
\newcommand\bbeta{\mbox{\boldmath $\beta$}}
\newcommand\bgamma{\mbox{\boldmath $\gamma$}}
\newcommand\bomega{\mbox{\boldmath $\omega$}}
\newcommand\blambda{\mbox{\boldmath $\lambda$}}
\newcommand\bmu{\mbox{\boldmath $\mu$}}
\newcommand\bphi{\mbox{\boldmath $\phi$}}
\newcommand\bzeta{\mbox{\boldmath $\zeta$}}
\newcommand\bsigma{\mbox{\boldmath $\sigma$}}
\newcommand\bepsilon{\mbox{\boldmath $\epsilon$}}

\newcommand{\be}{\begin{eqnarray}}
\newcommand{\ee}{\end{eqnarray}}
\newcommand{\nn}{\nonumber}

\newcommand{\ft}[2]{{\textstyle\frac{#1}{#2}}}
\newcommand{\eqn}[1]{(\ref{#1})}
\newcommand{\vsone}{\vspace{1cm}}
 
\begin{titlepage}

\begin{flushright}
hep-th/9907098\\
KCL-TH-99-26\\
\end{flushright}
\begin{centering}
\vspace{.2in}
{\Large {\bf Kinky D-Strings}}\\
\vspace{.4in}
Neil D. Lambert and David Tong${}$ \\
\vspace{.4in}
Department of Mathematics\\
Kings College, \\
The Strand, London,\\ 
WC2R 2LS, UK\\
\vspace{.05in}
{\tt lambert,tong@mth.kcl.ac.uk}\\
\vspace{.6in}
{\bf Abstract} \\
\end{centering}
\vspace{0.2in}
We study two-dimensional SQED viewed as the world-volume theory of a 
D-string in the presence of 
D5-branes with non-zero background fields that induce 
attractive forces between the branes. In various 
approximations, the low-energy dynamics is given by a 
hyperK\"ahler, or hyperK\"ahler with torsion, massive sigma-model. 
We demonstrate the existence of kink solutions corresponding to 
the string interpolating between different D5-branes. Bound 
states of the D-string with fundamental strings are identified 
with Q-kinks which, in turn, are identified with dyonic instanton 
strings on the D5-brane world-volume.
\vspace{.1in}


\end{titlepage}

\section{Introduction}

Supersymmetric gauge theories with eight supercharges generically 
posses a classical moduli space of vacua. Moreover 
non-renormalisation theorems 
prohibit the dynamical generation of a potential on this space 
by either perturbative or non-perturbative effects and 
the moduli space survives in the full quantum 
theory, albeit possibly differing from the classical 
space in its metric and singularity structure. Indeed, the 
existence of such quantum moduli spaces has been of 
paramount importance in determining many properties of the 
low-energy dynamics of theories with eight supercharges in two, 
three and four dimensions. 
\paragraph{}
However there are situations where, despite the existence of eight 
supercharges, the classical theory has only isolated vacua. Part of 
the motivation of the present paper is to investigate to what extent 
the low-energy dynamics of these theories can be described in terms of a 
potential on a quantum vacuum moduli space. We consider the simplest 
such model: two dimensional ${\cal N}=(4,4)$ 
SQED. As will be reviewed in section 2, the introduction of both 
Fayet-Iliopoulos (FI) parameters and hypermultiplet masses leads to 
a situation with only isolated vacua and, in different approximations, 
the low-energy dynamics is described as a massive sigma-model on 
different branches of the vacuum moduli space.
\paragraph{}
Further motivation comes from string theory where there exist 
brane configurations preserving eight supercharges that again 
have only isolated vacua. One such 
situation in type IIB theory has been described recently by 
Bergshoeff and Townsend \cite{bt}. 
These authors consider a $(1,1)$-string (i.e. a bound state of a D-string 
with a  fundamental (F-)string) lying parallel to $k$ separated D5-branes. 
As is well known, the D1-D5 system preserves $1/4$ of the spacetime 
supersymmetry, implying no force between a D-string and 
D5-brane. However, the F-string, and therefore the $(1,1)$-string under 
consideration, is attracted to the D5-branes. The result is a 
situation with $k$ isolated vacua corresponding to each of the 
possible $(1,1)$-string/D5-brane bound states. Moreover, in each 
of these vacua eight supercharges are again preserved. Bergshoeff 
and Townsend further showed that there should exist stable, BPS, 
kink configurations in which the $(1,1)$-string interpolates 
between two D5-branes as shown in figure 1. These solutions were 
identified with the T-duals of Q-kinks \cite{at,hl}.

\begin{figure}
\begin{center}
\epsfxsize=2.5in\leavevmode\epsfbox{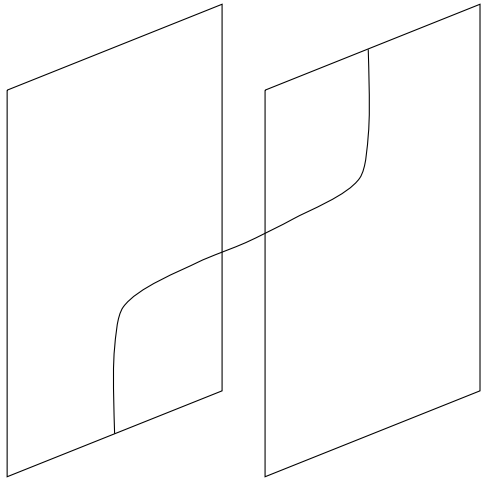}
\end{center}
\caption{\small{
The kinky string: such soliton solutions exist whenever the string 
forms a bound state with the D5-branes. This can be acheived either 
by considering a $(1,1)$-string, or by considering a D-string in the 
presence of constant background RR scalar or NS-NS two form.}}
\end{figure}

\paragraph{}
There exist related scenarios which capture the same physics. 
For instance, consider the two-dimensional $U(1)$ world-volume gauge 
theory of a D-string. The $(1,1)$-string bound-state corresponds 
to the introduction of a single quantised unit of field strength 
\cite{witbound}. From this perspective it is the non-zero world-volume 
electric field which leads to an attraction between the string and 
D5-branes. However one need not consider a full quantum of electric field. 
It was shown many years ago that the appearance of an unquantised, constant, 
electric field in two-dimensions can be interpreted as the addition of 
a $\theta$-angle to the Lagrangian \cite{col}. This in turn corresponds 
to the D1-D5 system in the presence of  a constant background 
Ramond-Ramond (RR) scalar field, $C_0$, as can be seen by considering 
the Wess-Zumino terms which couple a flat Dp-brane to the bulk RR potential, 
$C=\sum_nC_{\mu_1\cdots\mu_n}{\rm d}x^{\mu_1}\wedge\cdots
\wedge{\rm d}x^{\mu_n}$
\be
S_{\rm {W.Z.}}=\int_{{\bf R}^{p+1}}  
C\wedge e^{\cal F}\ .
\label{wz}\ee
Here ${\cal F}= F- B/2\pi\alpha^\prime$ where $F$
is the gauge field strength, $B$ is the pull back of the 
bulk NS-NS two-form potential, and $\alpha^\prime$ is the inverse 
string tension. This situation will prove somewhat easier to 
discuss from the gauge theory point of view and we will show in section 
3 that it does indeed lead to an attractive force between the string 
and D5-branes as claimed.
\paragraph{}
In fact there is a second way to attract a D-string and D5-brane: 
one may turn on a constant magnetic field on the D5-brane world-volume 
in directions orthogonal to the string. Let the D-string have world-volume 
$x^0, x^1$ and the D5-brane have world-volume $x^0,\cdots ,x^5$. 
We will ultimately be interested in turning on a constant 
self-dual field strength ${\cal F}_{ij}$, $i,j=2,3,4,5$, on the D5-brane 
(see next section) and vanishing field strength on the D-string. In 
order to see how this leads to a force between the string and five-brane, 
consider 
the bosonic terms in the action for an F-string stretched between them,
\be
S = {1\over 4\pi\alpha'}\int d\sigma d\tau\ \left\{
\eta^{\alpha\beta}\eta_{\mu\nu}\partial_\alpha X^{\mu}\partial_\beta X^{\nu} 
+ \epsilon^{\alpha\beta}B_{\mu\nu}\partial_\alpha X^{\mu}
\partial_\beta X^{\nu}\right\}
+ \int d\tau A_{\mu}{dX^\mu\over d\tau}\ ,
\label{openstringact}
\ee
where $\sigma \in [0,\pi]$ is the spatial coordinate of the open string
and the second term is evaluated at the boundary consisting of the 
two end points on the D-string ($\sigma=0$) and the D5-brane 
($\sigma=\pi$). 
The standard free field equations of motion are obtained from varying the
action \eqn{openstringact}. However the resulting boundary term 
for the bosons is modified to
\be
{1\over 2\pi\alpha'}\int d\tau\  \delta X^{\mu}\left(
\partial_\sigma X^\nu +
2\pi\alpha'{\cal F}^{\nu}_{\ \rho}\partial_\tau  X^\rho\right)\eta_{\mu\nu}\ .
\ee
Thus while Dirichlet (D) 
boundary conditions ($\delta X^\mu=0$) are consistent, 
we no longer have pure Neumann (N) boundary 
conditions but rather
\be
\partial_\sigma X^\mu +2\pi\alpha'{\cal F}^{\mu}_{\ \nu}\partial_\tau  X^\nu 
= 0\ .
\ee
For our current situation, only fields with DN boundary conditions are 
affected, namely $X^2, X^3, X^4, X^5$ and their fermionic partners.  
It is a staightforward exercise to see that
the modings of the fields are now shifted from the standard half-integer
moding for the bosons (and the corresponding integer or half-integer
moding for the fermions in the NS and R sectors respectively). We now
find two bosonic fields with moding $-\lambda + n$ and two bosonic
fields with moding $\lambda + n$. Here $n \in {\bf Z}$ and
\be
{\rm tan}^2(\lambda\pi) = {1\over \pi^2\alpha'^2{\cal F}^2} \ ,
\ee
where $0< \lambda \le 1/2$.
Similarly the fermions in the NS and R sectors are shifted from these
modings by $1/2$ and $0$ respectively. In particular, this 
leads to a tachyonic NS ground state from the DN sector, and the 
one loop open string amplitude for the potential between the 
D-branes \cite{dbrane} no longer enjoys the cancellation between the NS and 
R sectors arising from Jacobi's abstruse identity, thus leading to an 
attractive force between the branes. In fact, this calculation is 
essentially the same as for the force between moving D-branes performed 
in \cite{lifs}.  Note that if the same field strength is introduced 
at both ends of an open string, then the moding is not altered and 
the various forces cancel. So in particular there is still no force between 
the D5-branes. 
\paragraph{}
In summary, a force between the D-string and D5-branes may be 
generated by turning on either of the constant background 
fields, $C_0$ or $B_{ij}$. As will be 
reviewed in the following section, both of these spacetime 
background fields have a simple interpretation as parameters 
of the D-string world-volume theory. In the remainder of the 
paper, we will examine the physics of this system. The following
section reviews the 
the ${\cal N}=(4,4)$ $U(1)$ gauge theory that describes the low-energy 
dynamics of the D-string. For certain parameters, the theory has 
only isolated vacua corresponding 
to D-string/D5-brane bound states described above. We further 
demonstrate the existence of BPS soliton solutions of the classical  
equations of motion, although unfortunately we are unable to solve the 
Bogomol'nyi equation in general for this case. In the remaining two sections 
we consider two different approximations in which the low-energy dynamics 
reduces to a massive supersymmetric sigma-model on the Coulomb and 
Higgs branches respectively. The former description is unfortunately 
rather sick as the D-string is forced down the throat of the five-brane 
metric where the approximation breaks down and the 
physics is badly understood. Nevertheless, we are 
able to solve for the kinky D-string solutions in this case. The Higgs branch 
description is better behaved. Moreover, in this approximation we  
find a three-way identification between D-string/F-string bound 
states, Q-kinks \cite{at} and dyonic instanton strings \cite{us}. 
We also give a T-dualised description 
of the Higgs branch where the Q-kink momenta are exchanged in 
favour of winding modes \cite{hl}.

\section{The Model}

We will be interested in the limit of infinite Planck 
mass to ensure the suppression of the kinetic terms for the bulk closed 
string fields. In addition, the limit of vanishing string length, 
$\alpha^\prime\rightarrow 0$, allows us to ignore the higher order 
Born-Infeld interactions, and the D-brane dynamics reduces to a gauge 
theory. The configuration of D-string and D5-branes described in the 
introduction breaks ten-dimensional Lorentz invariance to,
\be
Spin(1,9)\rightarrow Spin(1,1)\times Spin(4)_R\times SU(2)_R\ ,
\label{decomp}
\ee
where $Spin(1,1)$ is the Lorentz group of the two-dimensional 
world-volume theory of the D-string, $Spin(4)_R$ describes the 
unbroken rotation group in $x^6,x^7,x^8,x^9$, transverse to 
the D5-branes, and $SU(2)_R$ describes self-dual rotations 
in the remaining directions tangent to the D5-brane, $x^2,x^3,
x^4,x^5$. The full $Spin(4)$ symmetry rotating these directions 
is not realised due to the orientation of the D-branes; had we 
considered anti-D5-branes, then the anti-self-dual rotations 
would have been realised. 
The D1-D5 system breaks $1/4$ of the spacetime supersymmetries, 
resulting in a ${\cal N}=(4,4)$ theory in two dimensions. 
\paragraph{}
The effective action for the D-string is determined by quantization 
of open strings with ends terminating on the D-string. Those that 
have both end points on the D-string yield a ${\cal N}=(4,4)$ 
vector multiplet, also known as a twisted multiplet, and a neutral 
${\cal N}=(4,4)$ hypermultiplet. Two complex scalars in the latter 
parametrise the position of the D-string in the $x^2,x^3,x^4,x^5$ plane. For 
a single D-string, these decouple and we shall ignore them for the 
remainder of the paper. The vector multiplet contains two further
complex, neutral, scalars, 
$\sigma$ and $\phi$, parametrising the position of the 
D-string in the directions $x^6,x^7,x^8,x^9$ transverse to the D5-branes. 
The superpartners of these scalars are a two-dimensional gauge potential, 
$A_\mu$, together with two Dirac fermions, $\lambda$ and $\chi$, 
which are uncharged under the gauge group. The vector mulitplet may be 
decomposed into an ${\cal N}=(2,2)$ gauge multiplet, $V$, and chiral 
multiplet, $\Phi$, with
\be
\{ A_\mu,\sigma,\lambda,D\}\in V\ \ \ \ {\rm and}\ \ \ \ \{\phi,\chi,F\}
\in \Phi\ ,
\nn\ee
where $D$ and $F$ are the usual real and complex auxiliary fields 
respectively. The field strength of $V$ is an ${\cal N}=(2,2)$ 
twisted chiral multiplet, 
$\Sigma =\bar{D}_+D_-V$, which has complex auxiliary field 
$D-iF_{01}$, where $F_{01}$ is the $U(1)$ field strength. 
Detailed conventions of ${\cal N}=(2,2)$ multiplets may be found in 
\cite{witten}. 
\paragraph{}
The presence of the D5-branes means that we must also
consider open strings with one end point on the D-string and the other on 
one of the $k$ D5-branes. These give rise to $k$ charged hypermultiplets, 
with the gauge coupling constant given by 
$e^2=g_s/\alpha^\prime$. Each of these hypermultiplets 
is composed of two ${\cal N}=(2,2)$ chiral 
multiplets, $Q_i$ and $\tilde{Q}_i$, $i=1,\cdots k$, each containing 
a complex scalar $q_i$ ($\tilde{q}_i$), a Dirac fermion, $\psi_i$ 
($\tilde{\psi}_i$) and a complex auxiliary field $F_i$ ($\tilde{F}_i$). 
All fields in $Q_i$ transform with charge $+1$ under the $U(1)$ gauge group, 
while those in  $\tilde{Q}_i$ transform with charge $-1$. 
\paragraph{}
The Lagrangian for $k$ massless hypermultiplets coupled to a $U(1)$ vector 
multiplet is given by ${\cal L}={\cal L}_{\rm D}+{\cal L}_{\rm F}$, 
where
\be
{\cal L}_{\rm D}&=&\int {\rm d}^4\theta\, \left\{\frac{1}{4e^2}
\left(\Phi^{\dagger}\Phi-\Sigma^\dagger\Sigma\right)
+\sum_{i=1}^k\left(\bar{Q}_i\,\exp(2V)
\,Q_i + \bar{\tilde{Q}}_i\, \exp(-2V)\,\tilde{Q}_i\right)\right\}\ ,
\label{LD}\ee
and
\be
{\cal L}_{\rm F}=\int {\rm d}^2\theta\, \left\{\sqrt{2}\sum_{i=1}^kQ_i
\Phi\tilde{Q}_i\right\}\ \ + \ \ {\rm h.c.}\ .
\label{LF1}\ee
The theory has a 
$H=Spin(4)_R\times SU(2)_R\times SU(k)$  global symmetry 
group, where the 
first two terms in the product are R-symmetries, and the latter is 
the flavour symmetry. The vector multiplet scalars, $\sigma$ and $\phi$,  
transform 
in the $({\bf 4}, {\bf 1}, {\bf 1})$ of $H$ while the hypermultiplet 
scalars, $q_i$ and $\tilde{q}_i$, transform as $({\bf 1},{\bf 3}+{\bf 1},{\bf 
k})$. 
\paragraph{}
There are further parameters that we may add to the Lagrangian. 
The existence of two complex mass parameters consistent with 
supersymmetry follows from the existence of the two complex scalars in the 
vector multiplet, each of which may induce a mass term for a hypermultiplet 
by the Higgs mechanism. From the string picture, the total mass 
(bare plus Higgs) of a 
hypermultiplet is determined by the distance from the D-string to 
the D5-brane, and the resulting mass parameters transform as 
$({\bf 4},{\bf 1},{\bf k}\otimes\bar{\bf k})$ under $H$. 
The complex matrix $m_{ij}$ appears in the Lagrangian as 
a hypermultiplet dependent vacuum expectation value (VEV) for 
$\phi$ and is referred to simply as the complex mass,
\be
{\cal L}_m=\int {\rm d}^2\theta\, \left\{\sqrt{2}\sum_{i,j=1}^k
m_{ij}Q_i\tilde{Q}_j\right\}\ \ +\ \ {\rm h.c.}\ .
\label{LM}\ee
We will work in a flavour basis in which the complex mass matrix is 
diagonal, $m_{ij}=m_i\delta_{ij}$ (no sum over $i$) and with 
$\sum_im_i=0$. 
The second mass parameter is equivalent to a hypermultiplet dependent 
VEV for $\sigma$, and is known as the twisted mass. In the 
diagonal flavour basis, it may be incorporated in the above 
Lagrangian by gauging the Cartan sub-algebra of the $SU(k)$ 
flavour symmetry in a ${\cal N}=(2,2)$ invariant fashion, thus 
introducing $k-1$ new gauge superfields, 
$V_i$, $i=1,...,k$, with $\sum_iV_i=0$, and with corresponding 
field strengths $\Sigma_i$. The hypermultiplet 
kinetic terms of \eqn{LD} are now given by the substitution,
\be
V\rightarrow V+V_i\ ,
\label{LD2}\ee
and a Lagrange multiplier is employed to restrict the complex scalar 
field that resides within $V_i$ to equal the twisted mass, denoted 
$\hat{m}_i$, 
\be
{\cal L}_{\rm L.M.}=\int{\rm d}^2\vartheta\  \left\{\frac{i}{2}\Lambda_i
(\Sigma_i-\hat{m}_i)\right\}\ \ +\ \ {\rm h.c.}\ ,
\label{LLM}\ee
where the measure ${\rm d}^2{\vartheta}$ denotes integration over the 
twisted half of superspace. Each Lagrange multiplier, $\Lambda_i$, 
is a twisted chiral superfield. These will play a prominent role 
in the T-duality of the Higgs branch discussed in section four. By 
construction, we have $\sum_i\hat{m}_i=0$.
\paragraph{}
Finally, two dimensional abelian gauge theories with eight supercharges 
also allow for the possibility of a dimensionless theta angle, 
$\theta$, and three dimensionless FI parameters, 
${\bzeta}=(\zeta_1,\zeta_2,\zeta_3)$. The former is a singlet under 
$H$, and we have already discussed its interpretation in the 
string theory: it corresponds to turning on a constant 
background RR scalar, as seen in \eqn{wz}. The FI parameters 
transform as $({\bf 1},{\bf 3}+{\bf 1},{\bf 1})$ under $H$. The 
FI parameters and theta-angle may be considered as vacuum expectation 
values of a background hypermultiplet. This fact, together with their 
transformation under $H$, is sufficient to identify their ten-dimensional 
spacetime interpretation and they correspond to a constant, background, 
self-dual NS-NS two form potential in the directions $x^2,x^3,x^4,x^5$ 
\cite{abs},
\be
\zeta_a\sim{\bf \eta}_a^{ij}{\cal F}_{ij}\ ,
\label{FIB}\ee 
where $\eta_a$ are the self-dual 't Hooft matrices. 
Both FI parameters and the theta angle may be incorporated in the 
Lagrangian as (twisted) F-terms,
\be
{\cal L}_{\rm F} =\int{\rm d}^2\theta\ W(\Phi)+\int{\rm d}^2{\vartheta}
\ {\cal W}(\Sigma)\ \ +\ \ {\rm h.c.}\ .
\label{FI}\ee
The superpotential 
$W(\Phi)=\hat{\tau}\Phi /2$ and the twisted superpotential 
${\cal W}=i\tau\Sigma /2$ depend upon the complexified combinations  
$\tau = i\zeta_3+\theta /2\pi$ and 
$\hat{\tau}=\zeta_1+i\zeta_2$. The effect of FI-parameters and theta 
angle on the D1/D5-system were also considered yesterday in a slightly 
different context \cite{dmw}. 
\paragraph{} 
We turn now to the vacuum moduli space of the theory. The classical 
potential energy, obtained by eliminating all auxiliary fields, is 
given by
\be
U&=&{e^2\over2}\left( \sum_{i=1}^k(|q_i|^2-|\tilde{q}_i|^2)-\zeta_3\right)^2
+{e^2\over2}\left(\sum_{i=1}^k (q^\dagger_i\tilde{q}^\dagger_i
+\tilde{q}_iq_i)-\zeta_1\right)^2 \label{vacuum}\\
&& +\, 
{e^2\over2}\left(i\sum_{i=1}^k(q^{\dagger}_i\tilde{q}^\dagger_i-\tilde{q}_iq_i)
-\zeta_2\right)^2 +2\sum_{i=1}^k\left(|\phi+m_i|^2
+|\sigma+\hat{m}_i|^2\right)\left(|q_i|^2+|\tilde{q}_i|^2\right)\ ,
\nn\ee
The structure of the classical vacuum moduli space, $U=0$, is 
dependent upon the values of the FI and mass parameters. 
We deal with each case in turn.
\paragraph{}
i) $m_i=\hat{m}_i=\bzeta=0$: \\
This case corresponds to zero background NS-NS two form flux and 
coincident D5-branes. There exist two branches of vacua:  
the Coulomb branch and the Higgs branch. The Coulomb branch has 
$q_i=\tilde{q_i}=0$, while the VEVs of $\sigma$ and $\phi$ are 
unconstrained, reflecting the fact that the D-string may roam 
the $x^6,x^7,x^8,x^9$ directions transverse to the D5-branes 
unimpeded. The metric on this space is the five-brane metric 
of \cite{chs} and will be reviewed in the following section. 
On the Higgs branch however, $\sigma ={\phi}=0$ while 
$q_i$ and $\tilde{q}_i$ are constrained 
only by the first three terms in \eqn{vacuum}. These constraints 
coincide with the ADHM equations for a single $U(k)$ instanton, 
resulting in a hyperK\"ahler quotient construction of a $4(k-1)$ 
dimensional space of vacua which coincides with the 1 instanton 
moduli space.
In the string theory interpretation, the D-string is 
absorbed by the D5-branes, where it appears as a single 
$U(k)$ instanton \cite{douglas}, as is apparant from \eqn{wz}. 
\paragraph{}
ii) $m_i\neq m_j$ or $\hat{m}_i\neq\hat{m}_j$ for $i\neq j$, and 
$\bzeta=0$: \\
This corresponds to the separation of the D5-branes. The Higgs 
branch is now lifted as a single D5-brane is unable to absorb a 
D-string (there are no finite action $U(1)$ instantons). In section 
4 we shall quantify the lifting of this moduli space for the 
simplest example of $k=2$. In fact, the lifting in more complicated cases, 
including multiple D5-branes and multiple D-strings,  
has been well understood for many years from the perspective of 
instanton calculus with spontaneously broken gauge groups. 
See for example \cite{inst}.
\paragraph{}
iii) $m_i=\hat{m}_i=0$, and $\bzeta \neq 0$: \\
The D5-branes remain coincident, but a non-zero constant background 
NS-NS two form flux is turned on \eqn{FIB}. The Higgs branch 
remains and coincides with the moduli space of a single $U(k)$ 
instanton on non-commutative ${\bf R}^4$ \cite{noncom}. The Coulomb 
branch is lifted, reflecting the 
attraction between the D-string and D5-branes as expected from 
the discussion in the introduction. In the following section, 
we quantify the lifting of the Coulomb branch.
\paragraph{}
iv)  $m_i\neq m_j$ or $\hat{m}_i\neq\hat{m}_j$ for $i\neq j$, and 
$\bzeta \neq 0$: \\
The D5-branes are separated, the NS-NS two form flux is turned 
on, and the D-string has only $k$ isolated vacuum states 
given by
\be
\phi=-m_i \ \ \ \;\ \ \ \ \sigma=-\hat{m}_i\ \ \ \ ;\ \ \ \ 
\eta_i\sigma_a\eta_i^\dagger=\zeta_a
\ \ \ \ \ \mbox{no sum over $i$}
\label{isovac}\ee
where $\sigma_a$ are the Pauli matrices and we have introduced the 
$SU(2)_R$ covariant vectors $\eta_i=(q_i , \tilde{q}_i^\dagger )$.  
We see from the first two equations above that each vacuum state 
occurs at the position of a D5-brane, corresponding to a 
D-string/D5-brane bound state.
\paragraph{} 
The theta angle has, of course, played no role in the above 
discussion. In the following section we shall integrate out 
all hypermultiplets, after which the $U(1)$ field strength, $F_{01}$, 
will play the role of an auxiliary field and we shall find that 
$\theta$ lifts the Coulomb branch in the same fashion as the FI 
parameters.
\paragraph{}
Let us now restrict attention to the fourth scenario above where, as 
discussed in the introduction, we may expect to find soliton solutions  
interpolating between two of the vacua \eqn{isovac}, 
corresponding to the eponymous kinky D-string. In order to simplify the 
equations, we consider the case $k=2$, and make full use of the 
$SU(2)_R\times Spin(4)_R$ R-symmetry to set
$\bzeta = (0,0,\zeta_3)$ (with $\zeta_3 >0$) and 
$m_1=m_2=0\ , {\hat m}_1 =-\hat{m}_2= i\mu$ (for real $\mu$). 
It is clear that a full
$SU(2)_R\times Spin(4)_R$ multiplet of BPS solitons must exist in 
the complete theory. Our search for solitons begins by requiring 
half of the $(4,4)$ supersymmeteries to be preserved. An additional 
simplification resulting from the above $SU(2)_R\times Spin(4)_R$ 
rotation is that now it is sufficient to search for
solutions which preserve half of the $(2,2)$ supersymmetry. In terms of 
${\cal N}=(2,2)$ superfields, these supersymmetry 
transformations take the form
\be 
\delta V &=& (\epsilon_+ Q_- + \epsilon_-Q_+)V \ ,\nn\\
\delta \Phi &=& (\epsilon_+ Q_- + \epsilon_-Q_+)\Phi \ ,\nn\\
\delta Q^i &=& (\epsilon_+ Q_- + \epsilon_-Q_+)Q^i \ ,\nn\\
\delta \tilde Q^i &=& (\epsilon_+ Q_- + \epsilon_-Q_+)\tilde Q^i \ .\nn\\
\ee
The component expansions for these expression can be found in \cite{witten}.
For the case in hand we find that, setting the fermion fields to zero,
supersymmetry is preserved if  $\epsilon_+=\epsilon_-,\ \eta_+=\eta_-$ and
the bososnic fields satisfy the first order equations
\be
\partial_x \sigma &=& {i\over\sqrt{2}} (D - iF_{01})\ ,\quad 
\partial_t\sigma=0\ ,\nn\\
\partial_x\phi &=&\partial_t \phi = F =  0\ ,\nn\\
D_x q^i &=& {i\over\sqrt{2}}(\sigma-\bar\sigma 
+2 \hat m^i)q^i
\ ,\quad 
D_tq^i = -{i\over \sqrt{2}}(\sigma+\bar\sigma)q^i\ ,\nn\\
D_x \tilde q^i &=&-{i\over \sqrt{2}}(\sigma-\bar\sigma 
+2\hat m^i)\tilde q^i\ ,\quad
D_t\tilde q^i = {i\over \sqrt{2}}(\sigma+\bar\sigma)\tilde q^i\ ,
\label{BPScon}
\ee
For $\zeta_3>0$, we see from \eqn{vacuum} that $\tilde{q}_i$ has 
vanishing VEV in both vacua and we may therefore trivially 
satisfy the last of these equations by $\tilde{q}_i=0$. We also find that,
although the Bogomol'nyi conditions \eqn{BPScon} admit solutions with 
a background electric
field $F_{01} = \partial_x(\sigma+\bar\sigma)/\sqrt{2}$,
the equations of motion require
$\sigma=-\bar\sigma$. Therefore we set $A_\mu=0$ and, 
after eliminating the auxiliary 
field $D$, we find the remaining coupled Bogomol'nyi equations
\be
\partial_x\sigma &=& -i{e^2\over\sqrt{2}}(|q^1|^2 + |q^2|^2 - \zeta_3)\ ,\nn\\
\partial_x q^1 &=& i\sqrt{2}q^1(\sigma + i\mu)\ ,\nn\\
\partial_x q^2 &=& i\sqrt{2}q^2(\sigma - i \mu)\ .
\nn\ee
The solutions  for the three functions $\sigma, q^1, q^2$ can now be written 
in terms of single function $\varphi=\varphi (x-x_0)$
\be
\sigma &=& -i\mu\frac{{\rm d}\varphi}{{\rm d}x} \ ,\nn\\
q^1 &=& \sqrt{\zeta_3}\exp\left(\sqrt{2}\mu [\varphi- (x-x_0)]\right)\ ,\nn\\
q^2 &=& \sqrt{\zeta_3}\exp\left(i\omega\right)
\exp\left(\sqrt{2}\mu [\varphi  + (x-x_0)]\right)\ , 
\label{sigqq}\ee
where $\varphi(x)$ itself satisfies the differential equation
\be
{{\rm d}^2\varphi \over {\rm d}x^2} = {e^2\zeta_3\over\sqrt{2}\mu}
\left[\exp\left(2\sqrt{2}\mu(\varphi-x)\right))+ 
\exp\left(2\sqrt{2}\mu(\varphi + x)\right) - 1\right]\ .
\label{soliton}\ee
This indeed describes a soliton solution interpolating between the first 
and second vacua as $x$ ranges from $-\infty$ to $+\infty$ provided 
$\varphi$ is assigned the boundary conditions
\be
\varphi(x)\rightarrow\mp x\ \ \ {\rm as}\ \ \ 
x\rightarrow\pm\infty
\nn\ee
Given these boundary conditions, there exists a unique solution for 
$\varphi$ and the soliton \eqn{soliton} posseses two collective coordinates. 
The first, $x_0$, describes the centre of mass of the kink. The second, 
$\omega$, has period $2\pi$ and describes the relative phase between 
the two vacua \footnote{The overall phase may be set to zero by a 
gauge rotation.}. 
\paragraph{}
We have been unable to solve equation \eqn{soliton} explicitly,
although for the special case 
$\mu^2=e^2\zeta_3/4$ we find
$\varphi(x)=-\frac{1}{\sqrt{2}\mu}\log (1+e^{2\sqrt{2}\mu x}) + x$. 
In general however, since the boundary conditions  
select a unique solution for $\varphi$,  we  
expect to find that the soliton solution has only the two zero modes $x_0$ and
$\omega$. The low energy effective dynamics of the soliton are then 
described by $N=4$ quantum mechanics with two bosonic fields.
\paragraph{}
Finally we consider the mass of the kink. The bosonic energy density for 
the fields $\sigma$, $q_1$ and $q_2$ is given by
\be
{\cal E} = {1\over e^2}|\partial_x \sigma -{i\over\sqrt{2}}D|^2 
+\sum_{i=1}^2 |\partial_iq^i -i\sqrt{2}q^i(\sigma +\hat m^i) |^2 + T\ ,
\ee
where
\be
T = {i\over\sqrt{2}e^2}D\partial_x\bar \sigma 
+ i \sqrt{2}\sum_{i=1}^{2}q^i(\sigma+\hat m^i)\partial_x \bar q^i + c.c.\ .
\ee
Notice that the first 
two terms are each positive definite and attain zero when the 
Bogomol'nyi equations are satisfied. 
The mass, $E$, of a Bogomol'nyi kink is therefore given by
\be
E = \int_{-\infty}^{\infty} T dx\ .
\ee
Substituting in the form \eqn{sigqq} for the solutions we find that
\be
T &=& \sqrt{2}\zeta_3\mu\left[\left({d^2\varphi\over dx^2}
+ 2\sqrt{2}\mu({d\varphi\over dx}+1)^2\right)
e^{2\sqrt{2}\mu(\varphi+x)}\right.\nn\\
&&\left.+\left({d^2\varphi\over dx^2}+ 2\sqrt{2}\mu
({d\varphi\over dx}-1)^2\right)
e^{2\sqrt{2}\mu(\varphi-x)} - {d^2\varphi\over dx^2}
\right]\ ,\nn\\
&=&{\zeta_3\over 2}{d^2 \over dx^2}\left[e^{2\sqrt{2}\mu(\varphi+x)}
+e^{2\sqrt{2}\mu(\varphi-x)} + 2\sqrt{2}\mu\varphi\right]\ ,
\ee
and we find $T$ to be a total derivative, with the rest 
mass of the kink  given by $E=2\sqrt{2}\zeta\mu$. This expression has a 
simple $SU(2)_R\times Spin(4)_R$ invariant extension, namely
\be
E=2M|\bzeta|
\label{massofthisthing}\ee
where $M=\ft 1{\sqrt{2}} (|m_1-m_2|^2+|\hat{m}_1-\hat{m}_2|^2)^{1/2}$ 
is the $Spin(4)_R$ invariant mass.

\section{On the Coulomb Branch}

In this and the following section, we consider the low-energy 
dynamics of the theory in different regions of the parameter 
space and discuss three further avatars of the kink solitons. 
One expects that at low-energies the physics is correctly 
described by a sigma-model on the classical vacuum moduli 
space. For $m_i=\hat{m}_i=\bzeta=\theta=0$ where we have both Coulomb and 
Higgs branches, consideration of the action of the R-symmetries 
on the scalars suggests that, despite strong coupling 
fluctuations, the Higgs and Coulomb branches decouple in the 
infra-red \cite{witstr,witcon}. Here we review the description of the 
Coulomb branch \cite{witcon, ds} 
and describe its lifting by the theta angle and FI parameters.
\paragraph{}
We consider first the situation of arbitrary masses, but 
with $\bzeta =0$ and $\theta =0$, ensuring the survival of the 
Coulomb branch. The 
classically massless superfields are the chiral field $\Phi$ and 
the twisted chiral field $\Sigma$. Up to two derivatives, the 
most general theory one can write down consistent with 
${\cal N}=(2,2)$ supersymmetry is
\be
{\cal L}=\int {\rm d}^4\theta\ K(\Phi ,\Phi^\dagger,\Sigma ,
\Sigma^\dagger).
\label{LC}\ee
$K$ is known as a generalised K\"ahler potential. In component 
form, the bososnic part of \eqn{LC} is given by a sigma-model 
with torsion
\be
{\cal L}_{\rm bose}&=&K_{\Phi^\dagger\Phi}(\partial_\mu\phi^\dagger
\partial^\mu\phi -F^\dagger F) - 
K_{\Sigma^\dagger\Sigma}(\partial_\mu\sigma^\dagger\partial^\mu
\sigma-\ft12 D^2-\ft12  F_{01}^2) \nn\\
&& +\, K_{\Phi\Sigma^\dagger}(\partial_\mu\phi\partial_\nu\sigma^\dagger)
\epsilon^{\mu\nu} 
+ K_{\Phi^\dagger\Sigma}(\partial_\mu\phi^\dagger\partial_\nu\sigma)
\epsilon^{\mu\nu}\ .
\label{Lbose}\ee
While \eqn{LC} is, by construction, invariant under ${\cal N}=(2,2)$ 
supersymmetry, further restrictions on $K$ are required in order 
for the Lagrangian to respect the full ${\cal N}=(4,4)$ algebra. 
If $K$ were a function of only chiral superfields, it is well 
known that it must give rise to a hyperK\"ahler metric. If however, 
as in the present case, $K$ is a function of both chiral and 
twisted chiral superfields, the condition on $K$ is \cite{ghr}
\be
K_{\Phi\Phi^\dagger}=-K_{\Sigma\Sigma^\dagger}\ .
\label{constraint}\ee
The resulting metric is not hyperK\"ahler but, rather, 
hyperK\"ahler with torsion. 
The constraint \eqn{constraint}, together with the requirement of 
$Spin(4)_R$ R-symmetry acting on the scalars $\sigma$ and $\phi$ 
is very restrictive and is sufficient to fix 
$K$ up to two constants, which are determined 
at tree level and one-loop \cite{arss,ds}. 
The resulting generalised K\"ahler potential is 
\be
K=\frac{1}{e^2}(\Phi^\dagger\Phi-\Sigma^\dagger\Sigma)
+\sum_{i=1}^k\left\{\log(\Phi+m_i)\log(\Phi^\dagger+m_i^\dagger)
-\int^{X_i}\frac{{\rm d}x}{x}\log (x+1)\right\}\ ,
\label{5bk}\ee
where the limit of the integral is given by the ratio
\be
X_i=\frac{(\Sigma+\hat{m}_i)(\Sigma^\dagger+\hat{m}_i^\dagger)}
{(\Phi+m_i)(\Phi^\dagger+m_i^\dagger)}.
\nn\ee
In the absence of a superpotential, the auxiliary fields, as well 
as the field strength, are set to zero by their equations 
of motion, and the bosonic action \eqn{Lbose} has the target space metric 
and torsion of $k$ five-branes \cite{chs}, with positions at 
$\sigma=-\hat{m}_i$ 
and $\phi=-m_i$, reflecting the fact the on the Coulomb branch the 
D-string probes the directions transverse to the D5-branes. 
The metric is given by 
\be
{\rm d}s^2=H(\phi,\phi^\dagger,\sigma,\sigma^\dagger)
\left({\rm d}\phi^\dagger{\rm d}\phi+{\rm d}\sigma^\dagger
{\rm d}\sigma\right)\ ,
\nn\ee
with
\be
H=\frac{1}{e^2}
+\sum_{i=1}^{k}\frac{1}{|\phi+m_i|^2+|\sigma+\hat{m}_i|^2}\ .
\nn\ee
As is well known, the five-brane metric has singularities at 
$\phi=-m_i$, $\sigma=-\hat{m}_i$, 
near which the metric has the form of an infinitely long tube. 
The existence of singularities on the Coulomb branch is of course 
familiar from examples in three and four dimensions and is 
usually indicative of a dual description of the physics. In the 
present situation, no dual description is known - see \cite{witcon} for 
a discussion of the meaning of the singularity. 
\paragraph{}
We turn now to the fate of the Coulomb branch with non-zero 
FI and theta parameters. These appear in the classical action as a 
superpotential term \eqn{FI}. The generalised K\"ahler potential is 
again fully determined at tree-level and one-loop to be \eqn{5bk}.
Note that there are no longer any sources for the gauge field $A_\mu$ so
we may treat $F_{01}$ as an auxiliary field. Now when we 
eliminate the auxiliary fields by their 
equations of motion we obtain  a sigma-model on the five-brane 
background with a potential, $V$, given by
\be
V(\phi,\phi^\dagger,\sigma,\sigma^\dagger)=\frac{1}{2}(\bzeta\cdot\bzeta
+\theta^2/4\pi^2)\, H^{-1}\ .
\label{pot}\ee
As expected, the potential has $k$ zeroes at the points $\phi=-m_i$ and 
$\sigma=-\hat{m}_i$, for each value of $i$, each corresponding to 
a D-string/D5-bound state. Moroever, the massive sigma-model with 
five-brane target space and potential \eqn{pot} is invariant under 
the full ${\cal N}=(4,4)$ supersymmerty algebra and all eight supercharges 
are preserved in each of the vacua. To see this, note 
that the five-brane metric admits two
sets of complex structures $(I^{\pm i}_{ j},J^{\pm i}_{ j},K^{\pm i}_{ j})$ 
which obey the algebra of
the quaterions and are covariantly constant with repect to the connection
with torsion $\Gamma^{\mp k}_{ij}$ \cite{chs}. 
Furthermore these complex structures are
in fact constant. From this, and using the criteria of \cite{pt}, it can be 
readily verified that this potential does indeed preserve the 
$(4,4)$ supersymmetry of the sigma-model.  
\paragraph{}
Notice that the FI paramaters and theta angle are present in 
\eqn{pot} in a $Spin(4)$ invariant fashion. Indeed, in \cite{witcon} 
it is argued that the $SU(2)_R$ R-symmetry under 
which $\bzeta$ transforms as a ${\bf 3}$ is enhanced in the 
infra-red to a second $Spin(4)$ R-symmetry. This effect was 
demonstrated in \cite{brodie} using IIA intersecting brane constructions 
of this theory, the extra dimension arising upon lifting to M-theory.  
\paragraph{}
Before progressing, it is important to determine in which 
limit of the theory the above description of massive vector 
multiplet fields is valid. In order to derive the potential \eqn{pot}, 
we have performed a one-loop calculation, expanding around 
configurations which are vacua {\em only} when the FI parameters 
vanish. One must check that the resulting description of the 
low-energy dynamics is consistent. Naively, we expect this to be 
the case for small FI parameters, $|\bzeta|\ll 1$. More quantatively, we 
require the potential energy of the Coulomb branch sigma-model to be 
less than the mass of each of the hypermultiplet fields that have 
been integrated out    
\be
\frac{1}{2}(\bzeta\cdot\bzeta+\theta^2/4\pi^2)H^{-1}\ll 
|\phi+m_i|^2+|\sigma+\hat{m}_i|^2-e^2|\bzeta| \ ,
\nn\ee
where the last term on the right hand side arises from the triplet of 
D-terms in the scalar potential \eqn{vacuum}. In fact, certain 
hypermultiplet fields that have been integrated out actually 
become tachyonic at a radius $e^2|\bzeta |$ from the five-brane 
singularity. Our conclusion is therefore that the massive Coulomb 
branch description of the low-energy dynamics is valid {\em except} 
in a region close to the vacua! However, despite this problem, 
we continue in our search for the kinky D-string soliton  
and are vindicated to some extent 
by the existence of a well behaved solution. 

In order to exhibit the existence of kink solitons in the 
Coulomb branch we can form a bound on the energy of a solution.
>From the Hamiltonian we find the energy of any configuration is given by
\be
E&=&\int {\rm d}x\,\left\{
H(\partial_t\phi\partial_t\phi^\dagger
+\partial_t\sigma\partial_t\sigma^\dagger)
+H(\partial_x\phi\partial_x\phi^\dagger
+\partial_x\sigma\partial_x\sigma^\dagger)
+\ft12(\bzeta\cdot\bzeta +\theta^2/4\pi^2)H^{-1}\right\}\ , \nn \\
&\geq& \int {\rm d}x\, \left\{ H\left| \partial_x\phi -
\gamma\frac{1}{\sqrt{2}}
(\bzeta\cdot\bzeta +\theta^2/4\pi^2)^{1/2}H^{-1}\right|^2
\right.\nn\\ 
&&\ \ \ \ \ \ \ \  +\ H\left|\partial_x\sigma -
\hat{\gamma}\frac{1}{\sqrt{2}}
(\bzeta\cdot\bzeta +\theta^2/4\pi^2)^{1/2}H^{-1}\right|^2 
\nn\\
&& \left.\ \ \ \ \ \ \ \ +\  \frac{1}{\sqrt{2}}(\bzeta\cdot\bzeta +
\theta^2/4\pi^2)^{1/2}\left(\gamma\partial_x\phi 
+\gamma^\dagger\partial_x\phi^\dagger+\hat{\gamma}\partial_x\sigma
+\hat{\gamma}^\dagger\partial_x\sigma^\dagger\right)\right\}\ , \nn\\
&\geq& \frac{1}{\sqrt{2}}(\bzeta\cdot\bzeta +\theta^2/4\pi^2)^{1/2}
\left.\left(\gamma\phi+\gamma^\dagger\phi^\dagger+\hat{\gamma}
\sigma+\hat{\gamma}^\dagger\sigma^\dagger\right)
\right|^{x=+\infty}_{x=-\infty}\ ,
\label{bound}\ee
where the first inequality is saturated by time independent configurations 
and where $\gamma$ and $\hat{\gamma}$ are both complex numbers satisfying 
$|\gamma|^2+|\hat{\gamma}|^2=1$. For solutions asymptoting 
to the two vacua
\be
\phi\rightarrow m_i \ \ {\rm and}\ \ \sigma\rightarrow \hat{m}_i
\ \ &{\rm as}&\ \ x\rightarrow\infty \ ,\nn\\
\phi\rightarrow m_j\ \ {\rm and}\ \ \sigma\rightarrow \hat{m}_j 
\ \ &{\rm as}&\ \ x\rightarrow -\infty\ ,
\nn\ee
the bound \eqn{bound} is maximised by 
\be
\gamma=\frac{m_i-m_j}{\sqrt{|m_i-m_j|^2+|\hat{m}_i-\hat{m}_j|^2}}
\ \ {\rm  and}\ \ 
\hat{\gamma}=\frac{\hat{m}_i-\hat{m}_j}
{\sqrt{|m_i-m_j|^2+|\hat{m}_i-\hat{m}_j|^2}}\ ,
\nn\ee
while the bound is saturated by solutions to the first order 
Bogomol'nyi equations
\be
\partial_t\left(\begin{array}{c} \phi \\ \sigma\end{array}\right)
=0\ ,\ \ \ \ \ \ \ \ \ 
\partial_x\left(\begin{array}{c}\phi \\ \sigma\end{array}\right)
=\frac{1}{\sqrt{2}}
\left(\begin{array}{c}\gamma \\ \hat{\gamma}\end{array}\right)
\left(\bzeta\cdot\bzeta+\theta^2/4\pi^2\right)^{1/2}
\, H^{-1}\ .
\nn\ee
For the simplest situation of two D5-branes, the kink solution 
describing a D-string interpolating between the five-branes is found 
to be given by $\phi=m_1\Pi(x)$ and $\sigma=\hat{m}_1\Pi(x)$, where 
$\Pi(x)$ satisfies the simple algebraic equation
\be
\left(\frac{1}{e^2}-\frac{4}{M^2}\frac{1}{\Pi(x)^2-1}
\right)\Pi(x)=\frac{\bzeta\cdot\bzeta +\theta^2/4\pi^2}
{M}(x-x_0)\ , 
\nn\ee
where, as in the previous section, the integration constant $x_0$ 
is the centre of mass of 
the kink. It is easy to check that the function on the left hand side is
monotonically increasing and can be inverted over the range $\Pi \in (-1,1)$,
$x \in (-\infty,\infty)$, yielding a previously unknown kink solution 
preserving $1/2$ of the 
supersymmetry. The energy of these solitons is easily determined from
\eqn{bound} to be
\be
E=2M\left(\bzeta\cdot\bzeta+\theta^2/4\pi^2\right)^{1/2}\ .
\label{Ec}
\ee
Notice that, with the exception of the contribution from 
the $\theta$-angle, the mass of this kink is the same as that 
of the classical soliton \eqn{massofthisthing}. The above  
Coulomb branch description includes quantum corrections however and 
the conclusion is that the masses of these states are not renormalised. 
This is in contrast to similar states in the ${\cal N}=(2,2)$ theories 
\cite{nick,ntd}.  
Curiously however, and unlike the solutions to the 
classical equations of motion described in the previous section, these 
solitons depend on only a single bosonic collective coordinate. The 
periodic collective coordinate describing the relative phase of the 
two vacua is missing from the above description. It appears that 
this situation has arisen because of the sickness of the model near 
the vacua.  As a check, one could consider the above $U(1)$ gauge 
theory with eight supercharges in three dimensions where 
the problems associated with the tube metric do not arise. 
In this case, the Coulomb branch is the $k$-centered Taub-NUT metric 
\cite{sw} with potential given by the length of the tri-holomorphic 
Killing vector. Solitons interpolating between two vacua are now 
strings in three dimensions and are in fact Q-kinks \cite{at}. Similar 
solitons will appear in the following section  and 
they do indeed have a second, periodic collective coordinate.

\section{On the Higgs Branch}

Our starting point in this section is the $4(k-1)$ dimensional 
Higgs branch which exists for coincident D5-branes. Fluctuations 
transverse to this space acquire a mass of order $e$ (the 
gauge coupling constant) and in the infra-red limit 
$e\rightarrow\infty$, the low-energy dynamics is well 
described a sigma-model on the Higgs branch. 
The metric on this branch receives no quantum corrections and 
arises as a hyperK\"ahler 
quotient of the $4k$ dimensional space parametrised by the 
hypermultiplet scalars, $q_i$ and $\tilde{q}_i$, with momentum  
maps given by the first three equations of the scalar 
potential \eqn{pot}. The three FI parameters correspond 
to the blow-up modes of the singularities of this space. 
In the simplest case of $k=2$ with the 
theta angle set to zero, the 
four-dimensional Higgs branch is the Eguchi-Hanson metric 
(for a derivation of this well-known fact see, for example, 
\cite{pol})
\be
{\rm d}s^2=G({\bf r}){\rm d}{\bf r}\cdot{\rm d}{\bf r}
+G({\bf r})^{-1}({\rm d}\psi-\bomega\cdot{\rm d}{\bf r})^2\ ,
\label{EH}\ee
with
\be
G({\bf r})=\frac{1}{|{\bf r}-\bzeta|}+\frac{1}{|{\bf r}|}
\ \ \ \ {\rm and}\ \ \ \ \nabla\times\bomega=\nabla G\ .
\label{G}\ee 
In terms of the hypermultiplet scalars, the coordinates on the metric 
\eqn{EH} are given by ${\bf r}=\eta_1\bsigma\eta_1^\dagger-\bzeta
=\eta_2\bsigma\eta_2^\dagger$ (where $\bsigma$ denotes the triplet of 
Pauli matrices and $\eta_i$ were defined after equation \eqn{isovac}) 
and  $\psi=2\, {\rm arg}\, (\tilde{q}_1^\dagger q_2^\dagger)$.
\paragraph{}
We now consider turning on mass terms for the hypermultiplets  
inducing a potential on the Higgs branch.
As in the previous section, the dictates of supersymmetry are strong 
enough to determine the form of the potential: it must be proportional to 
the length of a tri-holomorphic Killing vector. An explicit derivation of 
the potential will be given later in this section. In the case 
of Eguchi-Hanson, the Killing vector is simply 
$\partial /\partial\psi$, and the potential is given by 
\be
V({\bf r},\psi)= M^2 G^{-1}\ .
\label{hpot}
\ee 
Once again, the description of the low-energy dynamics 
in terms of such a model is valid only if the surviving modes are of 
lower energy than those that have been integrated out. For the 
vector multiplet this requires
\be
e^2(|{\bf r}|+|{\bf r}-\bzeta |)\gg M^2G^{-1}\ ,
\nn\ee
which is acheivable at all points of the Higgs branch providing 
$e\gg M$.  
This description therefore suffers from none of the sickness of the Coulomb 
branch. 
The low-energy dynamics is now described by a massive sigma-model with 
two isolated vacua at 
the fixed points of the isometry, ${\bf r}=0$ and ${\bf r}=\bzeta$. 
Once again, we expect to find soliton solutions interpolating 
between these two vacua. In fact, the properties of these solitons 
have been previously explored by Abraham and Townsend \cite{at}, 
where they were christened Q-kinks. As we now review, they have 
a rather unusual property for kinks in 2-dimensions: they are 
dyonic. That is, they have an internal degree of freedom which may 
be excited, resulting in a tower of kink-states, analagous to the 
tower of dyons that arises when quantising four-dimensional 
monopole configurations. In fact, 
it has been shown that for such dyonic kinks 
in ${\cal N}=(2,2)$ models, the similarity with four-dimensional 
dyons extends to the bound states and renormalised masses of 
these objects \cite{nick,ntd}. 
The energy of any classical configuration is given by
\be
E&=&\int {\rm d}x\,\left\{ (\dot{\bf r}\cdot
\dot{\bf r}+{\bf r}^\prime\cdot{\bf r}^\prime)G+ 
((\dot{\psi}+\bomega\cdot\dot{\bf r})^2 \right. \nn\\ 
&&\left.\ \ \ \ \ \ \ \ \ \ \ \ \ +\ (\psi^\prime+\bomega 
\cdot{\bf r}^\prime)^2)G^{-1} +M^2G^{-1}\right\}\ , 
\nn\ee
where the dot and prime denote temporal and 
spatial derivatives respectively. 
Following \cite{at}, we further introduce 
a unit four-vector, $(n_0,{\bf n})$, such that $n_0^2+{\bf n}\cdot{\bf n}=1$, 
and rewrite the energy by completing the square
\be 
E&=&\int {\rm d}x\  \left\{G\dot{\bf r}\cdot\dot{\bf r}+G^{-1}(\dot{\psi}
+\bomega\cdot\dot{\bf r}-Mn_0)^2 \right.\nn\\ 
&& \left. \hspace{.6in} + \ G^{-1}(\phi^\prime+\bomega\cdot{\bf r}^\prime )^2 
+G\left({\bf r}^\prime-G^{-1}M{\bf n}\right)^2 \right. \nn\\ 
&&\left.\hspace{.6in} + 2\left( G^{-1}(\dot{\psi}+\bomega\cdot\dot{\bf r})n_0
+{\bf r}^\prime\cdot{\bf n}\right)M\right\}\ , \nn\\
&\geq& 2M\left\{ Q_0n_0+{\bf Q}
\cdot{\bf n}\right\}\ .
\nn\ee
The final expression for the energy bound contains only conserved 
quantities, namely the Noether charge, $Q_0$, and 
topological charge, ${\bf Q}$, defined by
\be
Q_0= \int {\rm d}x\ G^{-1}\dot{\psi}\ ,\ \ \ \ \ \ \ 
{\bf Q}=\int {\rm d}x\ {\bf r}^\prime\ .
\nn\ee
The bound is clearly maximised by choosing $(n_0,{\bf n})
\sim(Q_0,{\bf Q})$, in which case we have
\be
E^2\geq 4M^2(Q_0^2+{\bf Q}\cdot{\bf Q})\ ,
\nn\ee
where the inequality is saturated by the Bogomol'nyi configurations 
satisfying, $\dot{\bf r}=0$ and $\psi^\prime=0$, together with
\be
\dot{\psi}&=&Mn_0\ ,  \nn\\ 
{\bf r}^\prime&=&M G^{-1}{\bf n}\ .
\nn\ee
The solutions to these equations, first found in \cite{at}, are 
given by 
\be
\psi&=&\psi_0 +Mn_0t\ , \nn\\
{\bf r}&=&\ft12 \bzeta\tanh \left(\ft12|{\bf n}| M
(x-x_0)\right)+\ft12\bzeta\ .
\label{qkink}\ee
There is thus a family of soliton solutions  parametrised by the 
angular velocity $n_0$, each with the two centre of mass collective 
coordinates, $\psi_0$ and $x_0$. The periodicity of $\psi_0$ ensures that 
upon quantisation the Noether charge $Q_0$ will be integer 
valued \cite{at}. What is the interpretation of the 
resulting tower of states in the ten-dimensional spacetime picture?  
We claim that they correspond to $(1,Q_0)$-strings (that 
is a bound state of the D-string with $Q_0$ F-strings) that 
interpolate between the two D5-branes in the manner described in 
the introduction. This clarifies the observation of \cite{bt} that 
such kinky string should have a description as Q-kinks. 
\paragraph{}
In order to 
elucidate this point, let us examine various properties of 
the Q-kinks. Firstly, we may consider the limit of vanishing 
FI parameters, $\bzeta=0$, in which the 
Eguchi-Hanson metric \eqn{EH} becomes the singular one-instanton 
moduli space. The 
potential now has only a single zero at the singular point ${\bf r}=0$ and 
the pure kink solution that carries no Noether charge shrinks 
to this point, reflecting the fact that the spontaneously 
broken gauge group that lives on the D5-branes cannot support 
a non-singular pure instanton solution. However, it was recently  
shown that non-singular ``dyonic'' instanton string solutions may exist in 
spontaneously broken gauge groups if the string also 
carries electric charge \cite{us}. Moreover, the description of these 
strings in terms of the instanton moduli space sigma-model is as 
a solution to the sigma-model equations of motion which 
coincides with the $\bzeta\rightarrow 0$ limit of 
the Q-kink solution \eqn{qkink}. Such strings break $1/4$ of the 
supersymmetry of the six-dimensional theory on the D5-branes and 
therefore $1/8$ of the 32 space-time supersymmetries. The 
only known string-like states with these properties are indeed the 
$(1,Q_0)$-strings interpolating between the two D5-branes. 
We note in passing that, using the results 
of \cite{noncom} and \cite{us}, the original Q-kink solution 
\eqn{qkink} with $\bzeta\neq 0$ describes a dyonic instanton 
string whose transverse space is non-commutative ${\bf R}^4$. 
Moreover, the existence of a such a soliton with zero electric charge  
reflects the fact that there exist smooth, non-dyonic, Abelian 
instantons in non-commutative spaces \cite{noncom}.
\paragraph{}
So far our discussion of the Higgs branch has been limited to zero 
theta angle. As shown in \cite{at} and \cite{nick}, the inclusion of 
$\theta$ induces a torsion on the Higgs branch sigma-model. 
To see this, consider the infra-red limit $e^2\rightarrow\infty$, in 
which the gauge field kinetic terms vanish, $A_\mu$ satisfies an 
algebraic equation of motion which may be substituted in the 
theta term
\be
S_\theta  = {\theta\over 2\pi}\int d^2 x\, F_{01} = 
{\theta\over 4\pi}\int d^2 x\,  
\epsilon^{\mu\nu}b_{IJ}\partial_\mu X^I\partial_\nu X^J\ ,
\ee
where $X^I = ({\bf r},\psi)$. By construction this term is a total
derivative and therefore does not affect the sigma-model equations of 
motion. However, it does affect the  theory 
through a shift in the Noether charge operator, in analogy with the 
Witten effect in four-dimensional gauge theories \cite{witeffect}.  
It is a straightforward task to adapt the
analysis presented in \cite{at} for the inclusion of $S_\theta$ to our case.
We find that the charge operator $Q_0=Q_\psi = -i\delta/\delta\psi$ 
is now shifted to
\be
Q_0 = Q_\psi  + {\theta\over 2\pi}\int_{-\infty}^{\infty}dx\  
b_{\psi a}\partial_x r^a\ . 
\ee
Using the relationship between the $(\bf r, \psi)$ coordinates of 
Eguchi-Hanson and the hypermultiplet scalars $(q^i,\tilde q^i)$ 
(given after equation \eqn{G}), we may determine $A_\mu$ 
from the original  action  to find
\be
b_{\psi a}\,\partial_xr^a  = 
{\partial_x}\left({|\bf r|\over |\bf r| + |\bf r -\bf \zeta|}\right)\ .
\ee
In this way we find for the Q-kink solitons that
\be
Q_0 = Q_\psi + {\theta\over 2\pi}\ .
\ee
Including this effect we recover, for $Q_\psi=0$, 
the same mass formula \eqn{Ec} 
that applies to the solitons on the Coulomb branch.
The periodicity of $\psi$ ensures that 
upon quantisation  $Q_\psi$ will be integer 
valued. The shift in the Noether charge of the Q-kinks that is 
induced by $\theta$ mirrors the effect of background the  
RR-scalar on the D-string/F-string bound states, where the allowed 
background electric field on the D-string is shifted from integral 
values \cite{witbound}. This supports our interpretation of Q-kinks 
with  $(1,Q_0)$-string bound states.
\paragraph{}
To find further evidence for this identification, 
let us consider how these states 
transform under T-duality, an operation that one can 
perform on any two-dimensional sigma-model with a $U(1)$ isometry. 
We use the ${\cal N}=(2,2)$ superfield duality transformations of 
Rocek and Verlinde \cite{rv}. T-dualisation of sigma-models with 
potentials was discussed in \cite{hl}, while application of these 
transformations to Higgs branches of theories with four supercharges 
were considered previously in \cite{ntd}. 
Our starting point is the microscopic Lagrangian defined in 
equations \eqn{LD}-\eqn{LLM}. 
The plan is to exchange each hypermulitplet, containing two 
${\cal N}=(2,2)$ chiral superfields of opposite charge, $Q_i$ and 
$\tilde{Q}_i$, for a ${\cal N}=(4,4)$ twisted multiplet, containing 
a single neutral ${\cal N}=(2,2)$ chiral multiplet 
$\Gamma_i=Q_i\tilde{Q}_i$, together with a neutral ${\cal N}=(2,2)$ 
twisted chiral multiplet, $\Lambda_i$, which is identified as the 
Lagrange multiplier introduced in \eqn{LLM}. We will only consider  
$\theta=0$ here and rewrite the FI parameter 
which appears as a twisted F-terms as the more usual D-term
\be
\frac{i}{2}\int{\rm d}^2\vartheta\ \Lambda_i\Sigma_i\ +\ 
{\rm h.c.}=i\int {\rm d}^4\theta\ \Lambda_iV_i\ +\ {\rm h.c.} \ .
\nn\ee
Using this trick, the full microscopic Lagrangian becomes 
${\cal L}={\cal L}_D+{\cal L}_F$, 
where ${\cal L}_D$ is given by \eqn{LD} together with the 
replacement \eqn{LD2} and the FI D-term
\be
-\frac{i}{2}\int{\rm d}^4\theta\ \sum_{i=1}^{k}
(\Lambda_i-\Lambda_i^\dagger)(V+V_i)\ ,
\nn\ee
while the F-terms are of the form \eqn{FI} with the (twisted) superpotentials 
given by
\be
W(\Phi,\Gamma_i)&=&\frac{i}{2}\hat{\tau}\Phi+\sum_{i=1}^k
\Gamma_i(\Phi+m_i) \ ,\nn\\
{\cal W}(\Sigma,\Lambda_i)&=&\frac{i}{2}\tau\Sigma-\sum_{i=1}^k
\ft12\Lambda_i(\Sigma+\hat{m}_i)\ .
\nn\ee
The above manipulations have led us to a reformulation of the 
microscopic action. This form is particularly useful for describing 
the Higgs branch soliton solutions. To this end, we first integrate 
out the gauge superfields, $V+V_i$. Moreover, in the 
strong coupling limit of the gauge theory, $e^2\rightarrow\infty$,  
the vector multiplet kinetic terms decouple and the fields $\Sigma$ 
and $\Phi$ become Lagrange multipliers and may also be integrated out, 
resulting in a ${\cal N}=(4,4)$ 
massive sigma-model, where the metric and torsion terms are given 
by ${\cal L}_{\rm D}=\int{\rm d}^4\theta\ {\cal K}$, with
\be
{\cal K}&=&\sum_{i=1}^k\left(-\ft14 (\Lambda_i-\Lambda_i^\dagger)^2
+\Gamma_i\Gamma_i^\dagger\right)^{1/2} \nn\\
&&+\frac{i}{2}\sum_{i=1}^k(\Lambda_i-\Lambda_i^\dagger)
\log\left[-\ft{i}{2} (\Lambda_i-\Lambda_i^\dagger)
+\left(-\ft14 (\Lambda_i-\Lambda_i^\dagger)^2+
\Gamma_i\Gamma_i^\dagger\right)^{1/2}\right]\ ,
\label{K}\ee
subject to the constraints arising from the elimination of the vector 
multiplet
\be
\sum_{i=1}^k\Lambda_i=\tau\ ,\ \ \ \ \ \ \ \ \ \sum_{i=1}^k\Gamma_i
=\hat{\tau}\ .
\label{hcons}\ee
While \eqn{K} leads to a Lagrangian manifestly invariant under 
${\cal N}=(2,2)$ supersymmetry, full ${\cal N}=(4,4)$ supersymmetry 
is preserved only if ${\cal K}$ satisfies \cite{ghr}
\be
\frac{\partial^2{\cal K}}{\partial\Gamma_i\partial\Gamma_j^\dagger}
=-\frac{\partial^2{\cal K}}{\partial\Lambda_i\partial\Lambda_j^\dagger}\ ,
\nn\ee
which indeed it does. The superpotentials are now simply 
$W=\sum_i\Gamma_im_i$ and ${\cal W}=-i\sum_i\Lambda_i\hat{m}_i/2$. 
\paragraph{}
Finally, we restrict attention once more to the case of $k=2$, where 
the constraints \eqn{hcons} may be easily solved, with $\Gamma=\Gamma_1
=\hat{\tau}-\Gamma_2$ and a similar expression for $\Lambda$. In order to 
exhibit the $SU(2)_R$ action on the Higgs branch, we introduce the 
3-vector superfield, ${\bf R}=({\rm Re}(\Gamma),{\rm Im}(\Gamma), 
{\rm Im}(\Lambda))$, together with the $SU(2)_R$ singlet, 
$\Xi={\rm Re}(\Lambda)$. If the scalar components are denoted using 
lower case version of their parent superfield, the T-dualised 
description of the Eguchi-Hanson Higgs branch has metric
\be
{\rm d}s^2=G({\bf r})({\rm d}{\bf r}\cdot{\rm d}{\bf r}+
{\rm d}\xi{\rm d}\xi )\ ,
\nn\ee
where $G({\bf r})$ is given once again by \eqn{G}. The model further 
differs from the original Higgs branch \eqn{EH} by a torsion term that may be 
easily derived from \eqn{K}. 
The potential on the T-dualised Higgs branch now 
arises from the superpotentials and is given by 
$V({\bf r})=M^2G({\bf r})^{-1}$. This is precisely the same as the
potential in the un-T-dualised theory \cite{hl}, thus providing an explicit 
derivation of \eqn{hpot}. 
The kink solitons in this model are now simply found using the 
techniques of the previous sections. As on the Coulomb branch, we 
insist upon time independent solutions to ensure the vanishing of the 
torsion contribution to the action. Once more, introducing a 
unit 4-vector, $(n_0,{\bf n})$, the energy of a time independent 
configuration is given by
\be
E&=&\int{\rm d}x\ \left\{G({\bf r})\left({\bf r}^\prime
\cdot{\bf r}^\prime+\xi^\prime\xi^\prime\right)
+M^2G({\bf r})^{-1}\right\} \ ,\nn\\
&=& \int{\rm d}x\ \left\{G\left({\bf r}^\prime-MG^{-1}{\bf n}\right)^2
+G\left(\xi^\prime-MG^{-1}n_0\right)^2\right\} \nn\\
&&\hspace{.7in}+2M{\bf r}^\prime\cdot{\bf n}+2M\xi^\prime n_0\ , \nn\\
&\geq& 2M\left.\left({\bf r}\cdot{\bf n}+\xi n_0\right)
\right|^{+\infty}_{-\infty}\ .
\nn\ee
In the familiar manner, the inequality is saturated for soliton solutions 
satisfying the Bogomol'nyi equations 
\be 
{\bf r}'=MG^{-1}{\bf n}\ \ \ \ {\rm and}\ \ \ \ \xi^\prime =MG^{-1}n_0\ ,
\nn\ee
Comparing with the Bogomol'nyi equations derived on the hyperK\"ahler 
Higgs branch, we find that $\xi^\prime=\dot{\psi}$. 
The Q-kinks with momentum in the T-dual direction are thus exchanged 
with winding configurations \cite{hl}. This provides further evidence for 
the identification of the Q-kink time dependence as fundamental 
strings. Finally, imposing the boundary conditions
\be
{\bf r}\rightarrow 0\ \ \ &{\rm and}&\ \ \ \xi\rightarrow 0\ \ \ {\rm as} 
\ \ \ x\rightarrow -\infty \ ,\nn\\
{\bf r}\rightarrow\bzeta\ \ \ &{\rm and}&\ \ \ \xi\rightarrow\Theta
\ \ \ {\rm as}\ \ \ x\rightarrow +\infty
\ .\nn\ee
for arbitrary $\Theta$. The energy bound is maximised by choosing 
$(n_0,{\bf n})\sim(\Theta,\bzeta)$, 
and the Bogomol'nyi equations are solved by
\be 
{\bf r}&=&\ft12\bzeta\tanh\left(\ft12|{\bf n}|M(x-x_0)\right)+\ft12\bzeta\ ,
\nn\\
\xi&=&\ft12\Theta\tanh\left(\ft12 n_0M(x-x_0)\right)+\ft12\Theta\ . \nn
\nn\ee
We again note that in the limit $\bzeta\rightarrow 0$, there still 
exist non-trivial solutions to the sigma-model equations of motion 
corresponding to D-string/F-string bound state kinks. 

\subsection*{Acknowledgements}

D.T. is supported by an EPSRC fellowship. We would like to thank 
Bobby Acharya, Harm Jan Boonstra, Nick Dorey, Jerome Gauntlett, 
Sunil Mukhi and Paul Townsend for useful  
discussions.

\end{document}